\documentclass{jpsj2}
\title{Lithium Ruthenates: Controlling Dimensionality and Topology of Magnetic-Ion Arrangements}

\author{Makoto \textsc{Soma}, and Hirohiko \textsc{Sato}\thanks{E-mail address:
hirohiko@phys.chuo-u.ac.jp}
}

\inst{Department of Physics, Chuo University, 1-13-27 Kasuga, Bunkyo-ku,
Tokyo 112-8551, Japan
 \\
}

\abst{
In order to control the dimensionality and the topology of the arrangement of magnetic Ru ions, NaCl-type lithium ruthenates have been investigated. Three types of lithium ruthenates, Li$_{3}$RuO$_{4}$, Li$_{2}$RuO$_{3}$, and cubic-Li$_{x}$RuO$_{1+x}$, were synthesized using a hydrothermal method. All of them have NaCl-type structure, but the arrangements of the cations, Li$^{+}$ and Ru$^{5+}$ (or Ru$^{4+}$), differ from each other. Li$_{3}$RuO$_{4}$ with one-dimensional zigzag chains of Ru ions undergoes an antiferromagnetic transition at 66 K and exhibits an irreversibile magnetism below 32 K, where the zero-field-cooled susceptibility differs from the field-cooled susceptibility. Li$_{2}$RuO$_{3}$ with a two-dimensional honeycomb network of Ru ions demonstrates a paramagnetism almost independent of temperature. A novel ruthenate cubic-Li$_{x}$RuO$_{1+x}$, in which Ru and Li randomly occupy the cation sites of NaCl lattice, undergoes a spin-glass transition at 10 K.
}

\kword{transition metal oxide, random structure, ruthenate, magnetism, crystal structure, low dimension, magnetism, spin glass, frustration
}

\begin{document}
\maketitle

\section{Introduction}
Among 4$d$ transition elements, ruthenium demonstrates outstanding properties in oxides. It is well known that a variety of electronic and magnetic phases appear in ruthenates. While the simplest binary oxide RuO$_{2}$ has a metallic conductivity with Pauli paramagnetism, perovskite-type SrRuO$_{3}$ exhibits a metallic conduction with itinerant ferromagnetism below 165 K, which is extraordinarily high for a $4d$ electron system.\cite{randall59,cao97} Especially, what has begun to arouse the wide interests of many physicists in ruthenates is the discovery of the superconductivity in Sr$_{2}$RuO$_{4}$.\cite{maeno94} Although the $T_{c}$ $(\approx 1 K)$ is not so high, many experiments such as NMR revealed its exotic spin-triplet properties.\cite{ishida98} On the other hand, there are many insulating ruthenates with localized spins.

Those spin systems with a singular geometry of magnetic ions are fascinating because of their non-trivial magnetic ground states.\cite{schollwock04} In one-dimensional antiferromagnets, development of a long-range N\'eel ordering is suppressed by a large thermal and quantum fluctuation. Furthermore, a gap state appears because of purely quantum mechanical reason if the spin number $S$ is integer. In case of two- or three-dimensional systems, antiferromagnetic triangular, kagom\'e or pyrochlore lattices hold a geometric frustration. In these systems, complicated phase transitions can occur in order to eliminate the frustration. Alternatively, the spins freeze in a spin-glass state without a long-range order. Spin glass is characterized by anomalous properties such as irreversible magnetism, frequency-dependent susceptibility and aging phenomena.\cite{binder86} Although, there are a number of experimental and theoretical studies on spin glass, many fundamental problems are still unsolved.

We are interested in searching new ruthenates with a various arrangements of magnetic Ru ions. In this study, we focused our interest on Li-Ru-O ternary systems with NaCl-type structure. In these ruthenates, Li$^{+}$ and Ru$^{n+}$ ($n=4$ or $5$) occupy the Na sites and O$^{2-}$ occupies the Cl sites. Because the ionic radius of Ru and that of Li are similar, their sites seem exchangable each other. We considered that this feature can be utilized for controlling topology, dimensionality and randomness of the Ru arrangement, which may cause a variety of exotic magnetisms. In many cases, high vapor pressure of lithium often prevents us from obtaining lithium-containing oxides by a solid-state reaction which is an ordinary method for a synthesis of oxides. Therefore, we used a hydrothermal method, by which we can keep high Li concentration in a solution at high temperature. As a result of many trials, we succeeded to synthesize several phases of NaCl type lithium ruthenates, one-dimensional Li$_{3}$RuO$_{4}$, two-dimensional Li$_{2}$RuO$_{3}$ and three-dimensional cubic-Li$_{x}$RuO$_{1+x}$. The last one is a novel compound and exhibits a spin-glass behavior.

\section{Experimental}
Lithium ruthenates were synthesized using a hydrothermal method. A mixture of Li$_{2}$O$_{2}$ (or Li$_{2}$O) and RuO$_{2}$ was sealed in a gold capsule with water or NaOH solution. Then it was put in a test-tube type autoclave. The autoclave was heated at 600-650 $^{\circ}$C in a furnace, keeping 150 MPa of hydrostatic pressure for a few days. The capacity of the capsule used for the hydrothermal synthesis was so small, about 1 ml, that the amount of obtained materials was only a few miligrams per batch. This prevented us from a chemical elementary analysis. An energy dispersive X-ray spectrometer (EDS) (Oxford, Inca Energy 500) installed on a scanning electron microscope was used just for checking if the sample includes Ru or Na. The crystal structures were characterized with a powder X-ray diffractometer. An imaging-plate type X-ray diffractometer (Rigaku, Rapid R-Axis) was also used when single crystals were obtained.

Magnetic susceptibilities were measured with a superconducting quantum-interference-device (SQUID) magnetometer (Quantum Design, MPMS-XL). For DC susceptibility measurements, powder samples were wrapped with a piece of aluminum foil and it was fixed in a straw. Samples for AC susceptibility were directly fixed on the inside of a straw with grease in order to avoid the effect of shielding current of aluminum. The specific heat was measured using an AC calorimeter. Powder sample was pressed into a disk-shaped pellet. 4 Hz of AC light pulses were irradiated on the face of the pellet fixed in a cryostat. The amplitude of the temperature oscillation measured by an AuFe-Chromel thermocouple fixed on the other face of the pellet was recorded using a lock-in amplifier. Specific heat is reciprocally proportional to the amplitude. We obtained only its temperature dependence without any estimations of the absolute value of the specific heat.

\section{Results and Discussion}
\subsection{Characterization and crystal structures}
\begin{figure}[tb]
\begin{center}
\includegraphics[width=1.0\linewidth]{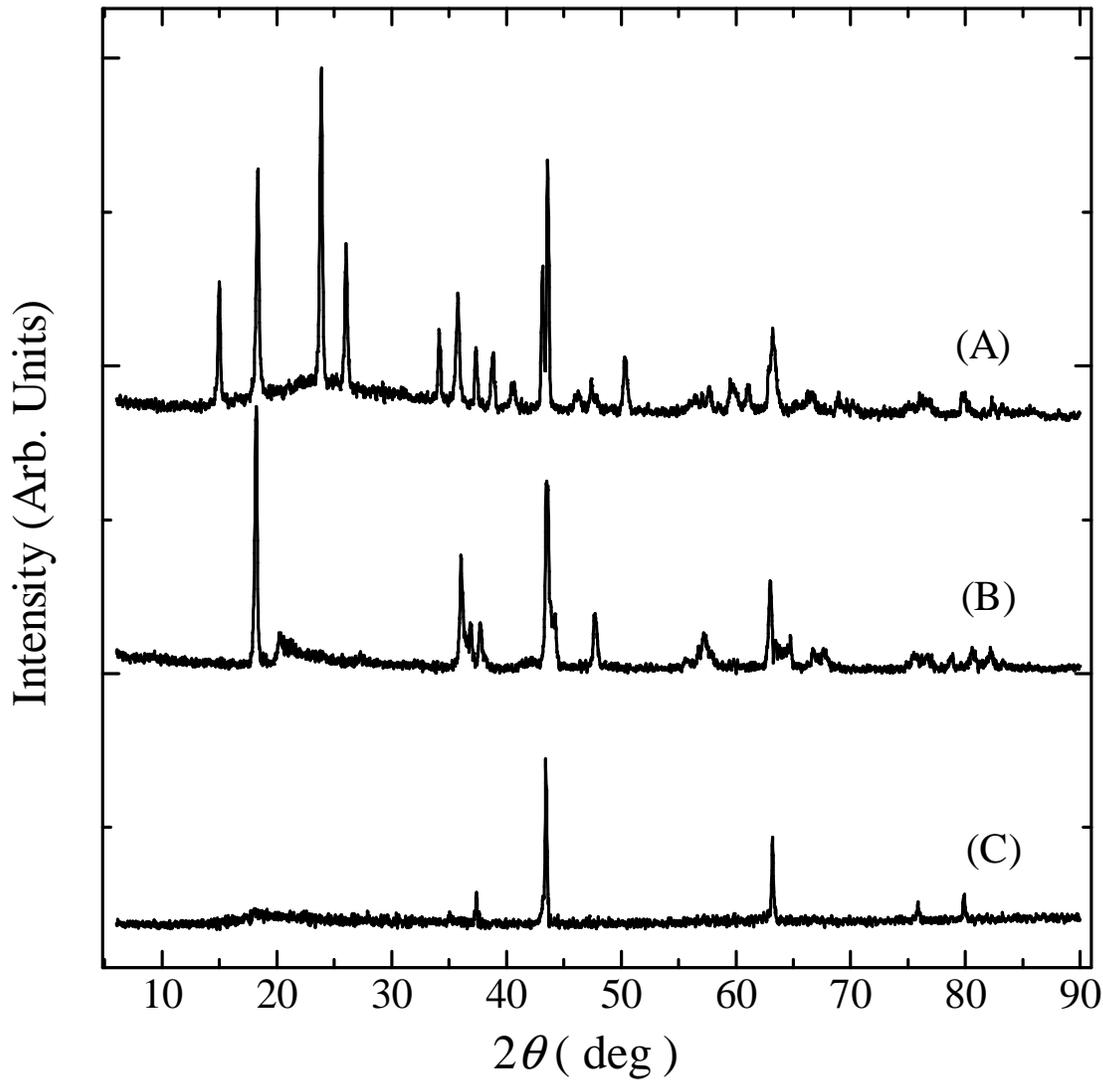}
\end{center}
\caption{X-ray powder diffraction patterns for the three lithium ruthenates. The symmetry and the lattice parameters are: (A)\emph{monoclinic}, $a = 5.13$ \AA$, b = 5.88$ \AA$, c = 5.10$ \AA$,\beta = 110.0^{\circ}$; (B)\emph{monoclinic}, $a = 4.92$ \AA$, b = 8.77$ \AA$, c = 9.88$ \AA$, \beta = 100.07^{\circ}$; (C)\emph{cubic}, $a = 4.144$ \AA.}
\label{fig1}
\end{figure}

Table~\ref{table1} summarizes the starting materials, the reaction conditions and the obtained materials. Depending on the synthetic condition, three types of materials, named (A), (B) and (C) were obtained. Figure~\ref{fig1} shows their X-ray powder diffraction patterns. The powder patterns of (A) and (B) almost agree with those of Li$_{3}$RuO$_{4}$\cite{alexander03} and Li$_{2}$RuO$_{3}$\cite{james88,kobayashi95}, respectively. Although they had been already synthesized by a solid-state reaction, the detailed structures are slightly different from ours as discussed below. On the other hand, the diffraction pattern of material (C) was indexed with a fcc lattice. No ruthenates with similar patterns have been, so far, reported. Therefore, (C) is a novel material firstly synthesized in this study and we named it cubic-Li$_{x}$RuO$_{1+x}$. 
\begin{table}
\caption{The conditions of the hydrothermal synthesis for the lithium ruthenates. The applied hydrothermal pressure was 150 MPa for all the syntheses.}
\label{table1}
\begin{tabular}{ccc}
\hline
Starting materials & $T$ ($^{\circ}$C) & Results\\
\cline{1-3}
RuO$_{2}$ (100 mg) + Li$_{2}$O$_{2}$ (100 mg) + H$_{2}$O (0.2 ml) & 650 &  (A) Li$_{3}$RuO$_{4}$\\
RuO$_{2}$ (100 mg) + Li$_{2}$O (100 mg) + H$_{2}$O (0.2 ml) & 650 &  (B) Li$_{2}$RuO$_{3}$\\
RuO$_{2}$ (35 mg) + Li$_{2}$O$_{2}$ (46 mg) + 4 M NaOH (0.4 ml) & 600 &  (C) cubic-Li$_{x}$RuO$_{1+x}$\\
\hline
\end{tabular}
\end{table}
We used a strong oxidant Li$_{2}$O$_{2}$ for the synthesis of (A), whereas Li$_{2}$O without oxidization power for (B). This resulted in Li$_{3}$RuO$_{4}$ with Ru$^{5+}$ and Li$_{2}$RuO$_{3}$ with Ru$^{4+}$, respectively. For material (C), we used Li$_{2}$O$_{2}$ and NaOH solution. The EDS analysis did not detect Na in the resulting material at all. This suggests that the role of NaOH was just controlling pH.

\begin{figure}[tb]
\begin{center}
\includegraphics[width=1.0\linewidth]{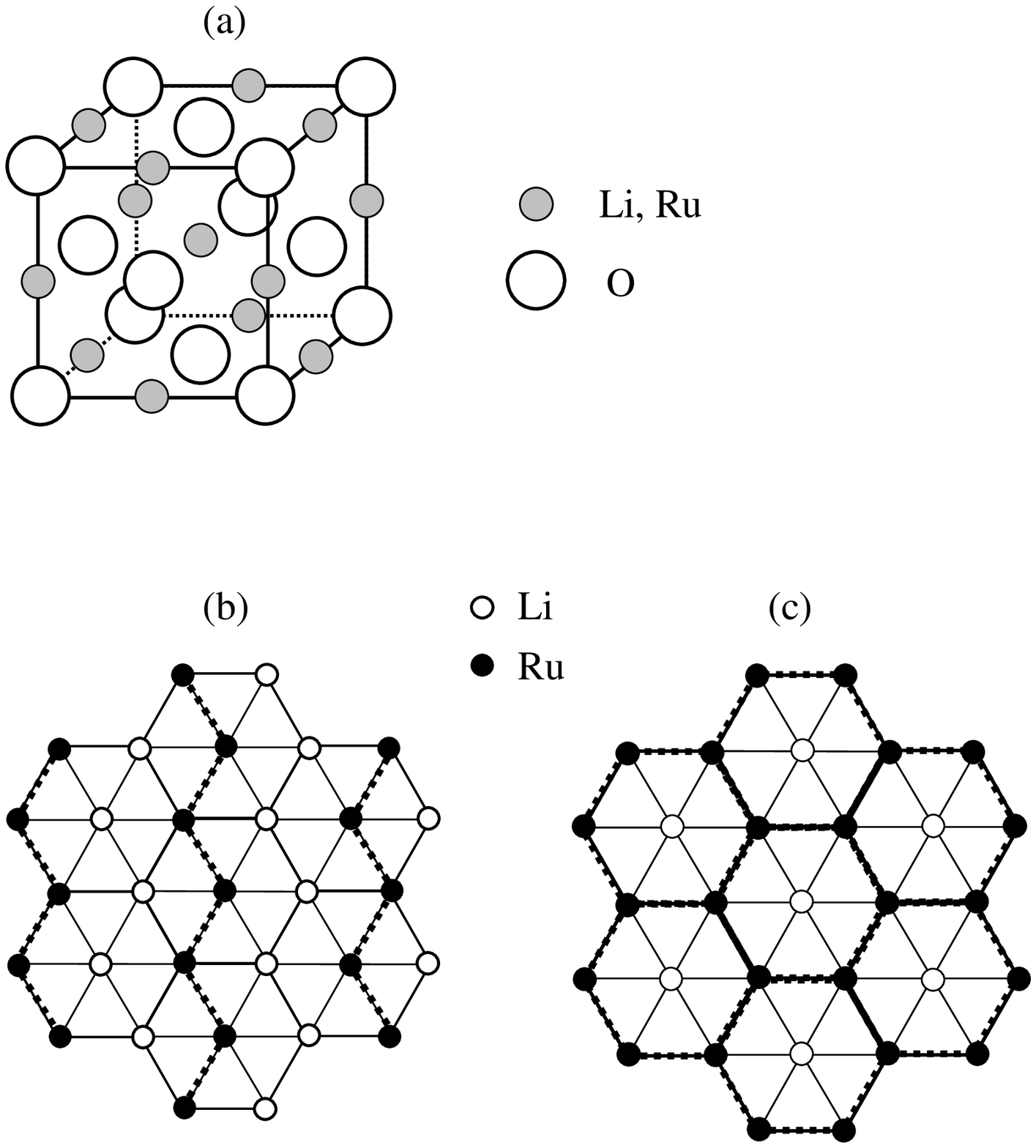}
\end{center}
\caption{Schematic views of the crystal structures of the lithium ruthenates. (a)NaCl-type lattice. Li and Ru occupy the octahedral interstitial sites of the fcc lattice of oxygens. (b)Cation arrangement of the LiRu-layer in Li$_{3}$RuO$_{4}$. Ru ions form one-dimensional zigzag chains. (c)Cation arrangement of the LiRu-layer in Li$_{2}$RuO$_{3}$. Ru ions form a two-dimensional honeycomb network.}
\label{fig2}
\end{figure}

Although the three phases have very different lattice parameters, thier structures are similar. Roughly speaking, all of them have NaCl-type structure. In other words, the structures are based on a fcc packing of oxygen. The cations, Ru$^{5+}$ (or Ru$^{4+}$) and Li$^{+}$, occupy the octahedral interstitial sites of the oxygen fcc lattice as shown in Fig.~\ref{fig2}(a). The variety of the cation arrangements makes these three phases different.
Li$_{3}$RuO$_{4}$ and Li$_{2}$RuO$_{3}$ are characterized by an alternating stacking of two types of cation layers, Li-layer and LiRu-layer, along the 111 direction of the cubic lattice. The Li-layer is a triangular lattice of lithium atoms and the LiRu-layer contains Li and Ru whose arrangement depends on materials as shown in Fig.~\ref{fig2}(b) and Fig.~\ref{fig2}(c). In Li$_{3}$RuO$_{4}$, Ru ions form one-dimensional zigzag chains which are separated each other by LiO$_{6}$ octahedra. On the other hand, Ru atoms form a two-dimensional honeycomb network in Li$_{2}$RuO$_{3}$.
The structures explained above are too idealized. Our X-ray powder patterns detected disorders. The result of a Rietveld analysis on Li$_{3}$RuO$_{4}$ using RIETAN 2000\cite{izumi00} is summerized in Table~\ref{table2}. We first tried to refine the structure assuming an ideal composition. This resulted in large $R_{wp}$. Then, we assumed that ruthenium atoms substitute for a part of lithium atoms. In the optimized structure, 7.7 \% of Li are replaced by Ru. Alexander, \emph{et al.} reported a similar result on Li$_{3}$RuO$_{4}$ obtained by a solid-state reaction. In their case, the amount of the substitution is only 1.4 \%, smaller than in our sample.
\begin{table}
\caption{Atomic positions and occupancies of Li$_{3}$RuO$_{4}$ optimized by a Rietveld analysis. The temperature factors $B_{eq}$ are fixed to 0.3. The final reliability factors are $R_{wp} = 0.098$ and $S=2.13$. }
\label{table2}
\begin{tabular}{cccccc}
\hline
Atom & $x$ & $y$ & $z$ & $B_{eq}$ & Occ.\\
\cline{1-6}
Li1/Ru1  & 0.0 &  0.645(4) &  0.25 &  0.3 & 0.93/0.07\\
Li2/Ru2  &  0.5 &  0.843(4) &  0.25 &  0.3 & 0.93/0.07\\
Li3/Ru3  & 0.5 &  0.397(2) &  0.25 &  0.3 & 0.90/0.10\\
Ru4  & 0.0 & 0.138(1) & 0.25 & 0.3 & 1.0\\
O1  & 0.233(2) & 0.116(2) & 0.002(2) & 0.3 & 1.0\\
O2  & 0.238(2) & 0.374(2) & 0.493(2) & 0.3 & 1.0\\
\hline
\end{tabular}
\end{table}
In the case of Li$_{2}$RuO$_{3}$, there was a significant difference between the powder pattern of the hydrothermal sample and that of the samples synthesized by a solid-state reaction following Ref.~\cite{james88}. A part of reflections was broad and weak in the hydrothermal sample. A quantitative analysis indicates that ruthenium atoms substitute for a part of lithium sites in the LiRu planes. 

\begin{figure}[tb]
\begin{center}
\includegraphics[width=1.0\linewidth]{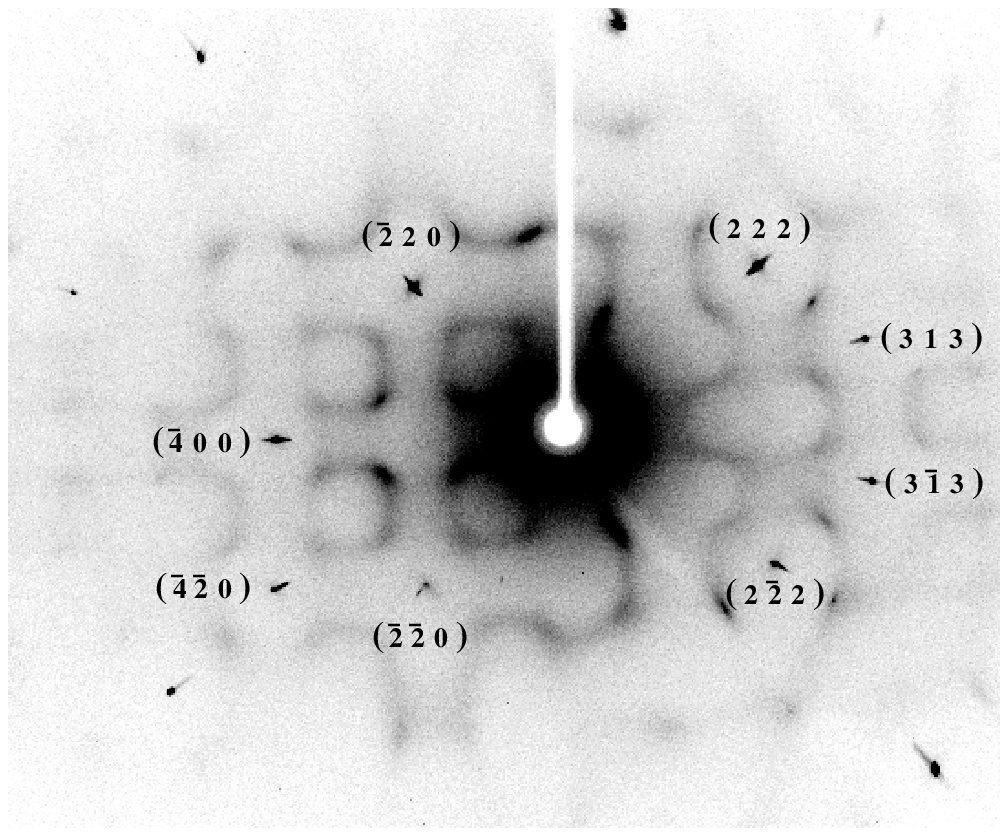}
\end{center}
\caption{X-ray oscillation photograph of cubic-Li$_{x}$RuO$_{1+x}$ single crystal. A Mo-K$\alpha$ beam was incident on the single crystal with an oscillation angle of 30$^{\circ}$. The scattered X-ray was recorded on a cylindrically-shaped imaging plate. In addition to sharp Bragg reflections whose indices are indicated, motley diffuse scatterings are observed. The scattering intensity at the streak is far weaker than that at Bragg spots although the contrast of the diffuse scatterings is emphasized. The white circle on the center with a line is the silhouette of a direct-beam catcher.}
\label{fig3}
\end{figure}

In cubic-Li$_{x}$RuO$_{1+x}$, the cation sites of the NaCl-type lattice are randomly occupied by Li and Ru. We also succeeded in obtaining single crystals of this phase. The typical size of the crystals was $100\times100\times100$ $\mu$m$^{3}$. One of the X-ray oscillation photographs on a single crystal is shown in Fig.~\ref{fig3}. The photograph is characterized by strong and sharp diffraction spots and streaky diffuse scatterings. The analysis using only the sharp spots gives a cubic unit cell with $a = 4.144$ \AA, which coincides the unit cell of the fcc oxygen lattice. This indicates that there is no long-range order of the cations located in the octahedral interstitial sites of the fcc lattice. From the intensity data, we tried to determine the composition of this material. We assumed that oxygen forms a perfect fcc lattice and that Ru and Li randomly occupy the octahedral sites. Then, their occupancies were optimized as parameters. The obtained Ru occupancy was about 0.25, whereas we did not succeeded in optimizing the Li content because it have a very small effect on the $R$ factor. Assuming no vacancies at cation sites, the nominal composition based on this analysis is closed to ``Li$_{3}$RuO$_{4}$''. This indicates that the oxidation number of Ru is 5+, which is consistent with the synthetic condition with a strong oxidant, Li$_{2}$O$_{2}$. The existence of the diffuse scatterings proves that the randomness in the cation arrangement is not perfect but that there is a short-range order. Their streaky shape indicates that the three-dimensional correlation of the short-range order is weak.
There are several batches which exhibited strong scatterings at $(l/2, m/2, n/2)$, where $l$, $m$ and $n$ are integers. In those batches, a long-range superstructure of the cations is developed. A similar type of cation ordering is also reported in LiFeO$_{2}$.\cite{anderson64} 

\subsection{Magnetism of Li$_{3}$RuO$_{4}$}
\begin{figure}[tb]
\begin{center}
\includegraphics[width=1.0\linewidth]{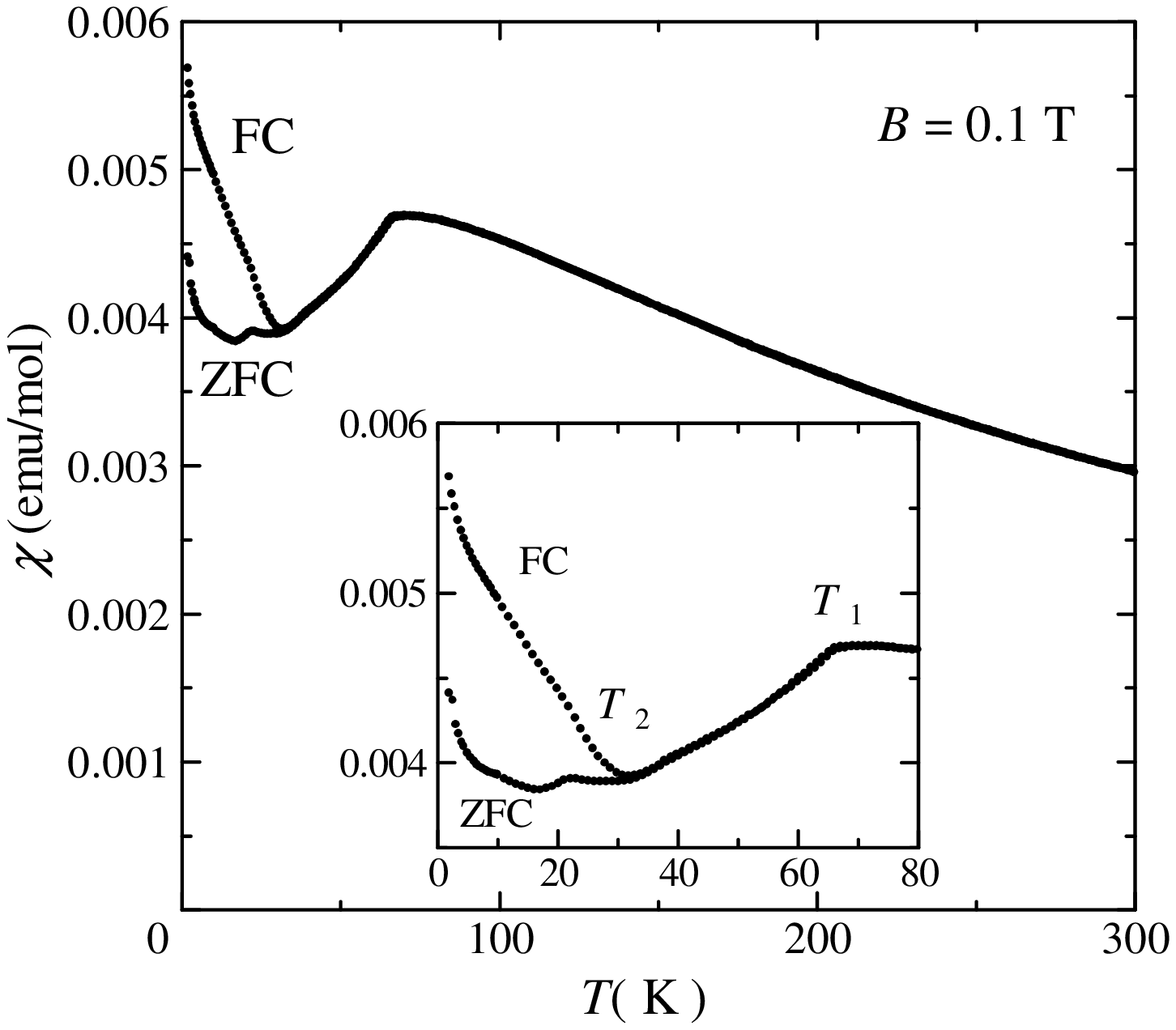}
\end{center}
\caption{Temperature dependence of the magnetic susceptibility of Li$_{3}$RuO$_{4}$. The susceptibility exhibits a sharp bend at $T_{1}$ and an irreversibility below $T_{2}$ as shown in the inset.}
\label{fig4}
\end{figure}

Figure~\ref{fig4} shows the temperature dependence of the magnetic susceptibility of polycrystalline Li$_{3}$RuO$_{4}$ under 0.1 T of magnetic field. The magnetic susceptibility exhibits a Curie-Weiss type behavior above 100 K. A local maximum with a sharp bend appears at $T_{1} = 66$ K, below which the susceptibility decreases with decreasing temperature down to $T_{2} (= 32$ K). Below $T_{2}$, a discrepancy between the field-cooled (FC) curve and the zero-field-cooled (ZFC) curve appears. The FC susceptibility steeply rises with decreasing temperature. The ZFC susceptibility exhibits a small hump at 22 K. We measured the susceptibility under several magnetic fields. The transition temperatures, $T_{1}$ and $T_{2}$, did not depend on magnetic fields, but the magnitude of the discrepancy below $T_{2}$ was more pronounced for lower magnetic fields.
We fitted the data above 100 K with Curie-Weiss formula. The obtained Curie constant and Weiss temperature were $C = 1.7$ emu$\cdot$K/mol and $\Theta = -231$ K, respectively. Since Ru$^{5+}$ has three 4$d$ electrons in the $t_{2g}$ orbitals, each Ru has $S=3/2$ spin. Assuming $g=2$, the Curie constant is calculated as 1.875 emu$\cdot$K/mol, in good agreement with the observation.

\begin{figure}[tb]
\begin{center}
\includegraphics[width=1.0\linewidth]{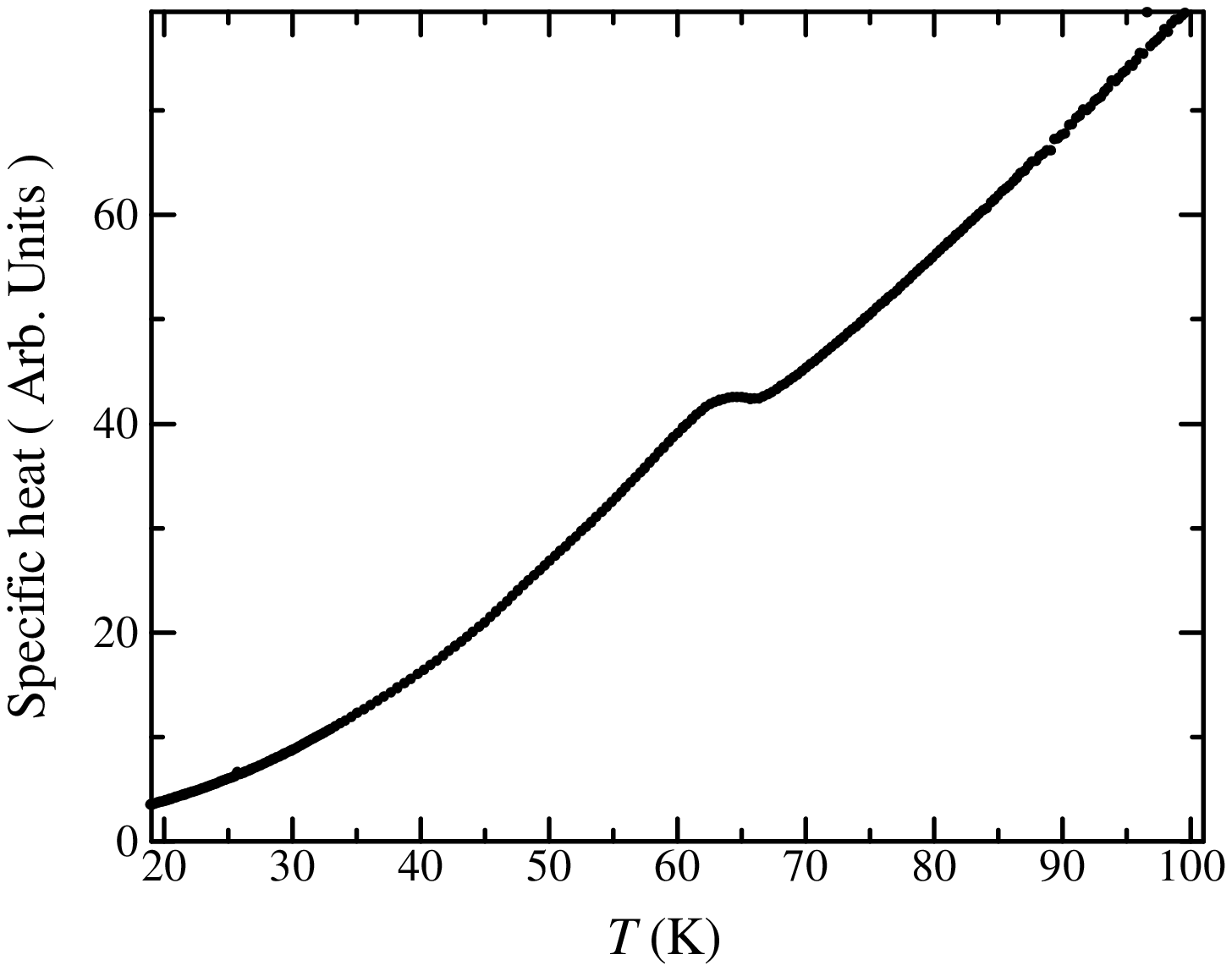}
\end{center}
\caption{Specific heat of Li$_{3}$RuO$_{4}$ measured with an AC caloriemeter. An anomaly is detected around at $T_{1}$.}
\label{fig5}
\end{figure}

Alexander, \emph{et al.} reported the magnetic susceptibility of Li$_{3}$RuO$_{4}$ prepared by a solid-state reaction.\cite{alexander03} Their data are similar to ours although their transition temperatures are lower than ours; their $T_{1}$ and $T_{2}$ are 50 K and 10 K, respectively. They also mentioned that a neutron diffraction study at 5 K did not detect any additional magnetic Bragg scatterings. Therefore, they interpreted that the anomaly at $T_{1}$ is due to a short-range order. However, the anomaly in the susceptibility at $T_{1}$ seems very sharp and clear, unlike a broad maximum coming from a development of a short-range order in one-dimensional antiferromagnet.\cite{bonner64} In order to clarify whether there is a phase transition at $T_{1}$ or not, we carried out a specific heat measurement. Figure~\ref{fig5} shows the temperature dependence of the specific heat. An anomaly is clearly observed at $T_{1}$, suggesting the existence of a phase transition at $T_{1}$. The peak at $T_{1}$ does not seem as sharp as that of ordinary second-order phase transition. This is probably because we used powder sample in spite of the AC method. In this case, thermal resistances between the particles often cause a problem. Although we have no information about the magnetic structure below $T_{1}$, a N\'eel type ordering shown in Fig.~\ref{fig7} does not conflict with their comment on the neutron result so much because the magnetic unit cell of this type ordering is the identical with  the structural unit cell. Such a magnetic order might be overlooked by the neutron measurement because it slightly changes the intensity of each Bragg reflection without occurrence of any extra peaks.

The $T_{1} / |\Theta| $ ratio is as small as 0.3. This means that something suppresses a development of the long-range antiferromagnetic order at the mean-field transition temperature. The one-dimensionality of the zigzag chain should be a dominant origin. The development of the three-dimensional long-range order below $T_{1}$ indicates that the interchain exchange interaction are not negligible. A super-superexchange interactions through LiO$_{6}$ octahedra might be strong enough to couple the chains. We should also remind that a part of interchain Li are substituted by Ru. They can, to some extent, mediate the interchain exchange interactions. The substitution ratio is higher in the hydrothermal sample than in the sample prepared by a solid-state reaction. This could explain that the hydrothermal sample has higher transition temperatures.

The irreversible behavior below $T_{2}$ indicates another magnetic transition. The appearance of an irreversibility far below a magnetic phase transition temperature reminds us of reentrant spin-glass phenomena. In Fe$_{x}$Mn$_{1-x}$TiO$_{3}$, for example, an irreversible magnetism appears below some temperature far lower than the antiferromagnetic transition temperature.\cite{ito88} There are, however, qualitative differences between typical reentrant spin-glass systems and our Li$_{3}$RuO$_{4}$. As shown in Fig.~\ref{fig4}, the FC curve of Li$_{3}$RuO$_{4}$ steeply increases below the onset temperature of the irreversibility, which is unlike the temperature dependence of FC susceptibility in reentrant spin glass systems. In general cases, a spin-glass system exhibits aging phenomena; the magnetization in a glass state becomes time dependent in a wide range of time scales. We measured the time dependence of the magnetization for several hours below $T_{2}$. However, the magnetization always keeped constant value within the resolution of our instrument.

\begin{figure}[tb]
\begin{center}
\includegraphics[width=1.0\linewidth]{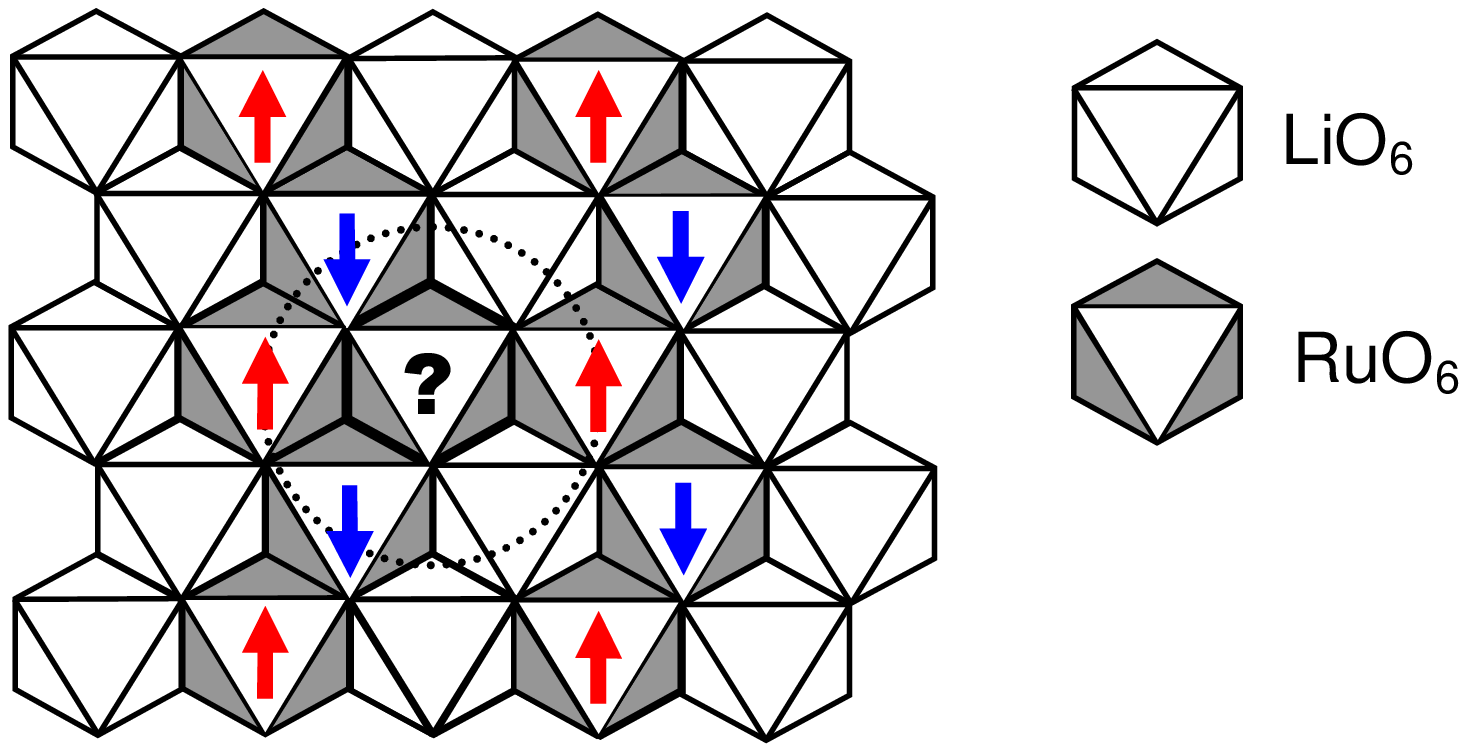}
\end{center}
\caption{Possible antiferromagnetic order below $T_{1}$ in Li$_{3}$RuO$_{4}$. A geometry around the interchain Ru site is also shown. The direction of the interchain spin is not determined because of the frustration.}
\label{fig6}
\end{figure}

Another scenario about the $T_{2}$ transition is as follows. The super-superexchange interaction via Li causes a N\'eel order of the magnetic moments located on the zigzag chains at $T_{1}$, but the magnetic moments on the interchain Ru atoms which partially substitute the Li sites are not necessarily ordered. This partially melted state could be caused by a geometric frustration. The interchain Ru atoms has four neighbors as shown in Fig.~\ref{fig6}. Assuming the same magnitudes of antiferromagnetic interactions, the directions of the spins of these Ru are not determined by nearest neighbor exchange interactions. The $T_{2}$ transition is possibly interpreted as a phase transition where these interchan spins order at last. The FC susceptibility suggests that the $T_{2}$ transition is accompanied by a weak ferromagnetism.

\subsection{Magnetism of Li$_{2}$RuO$_{3}$}
\begin{figure}[tb]
\begin{center}
\includegraphics[width=1.0\linewidth]{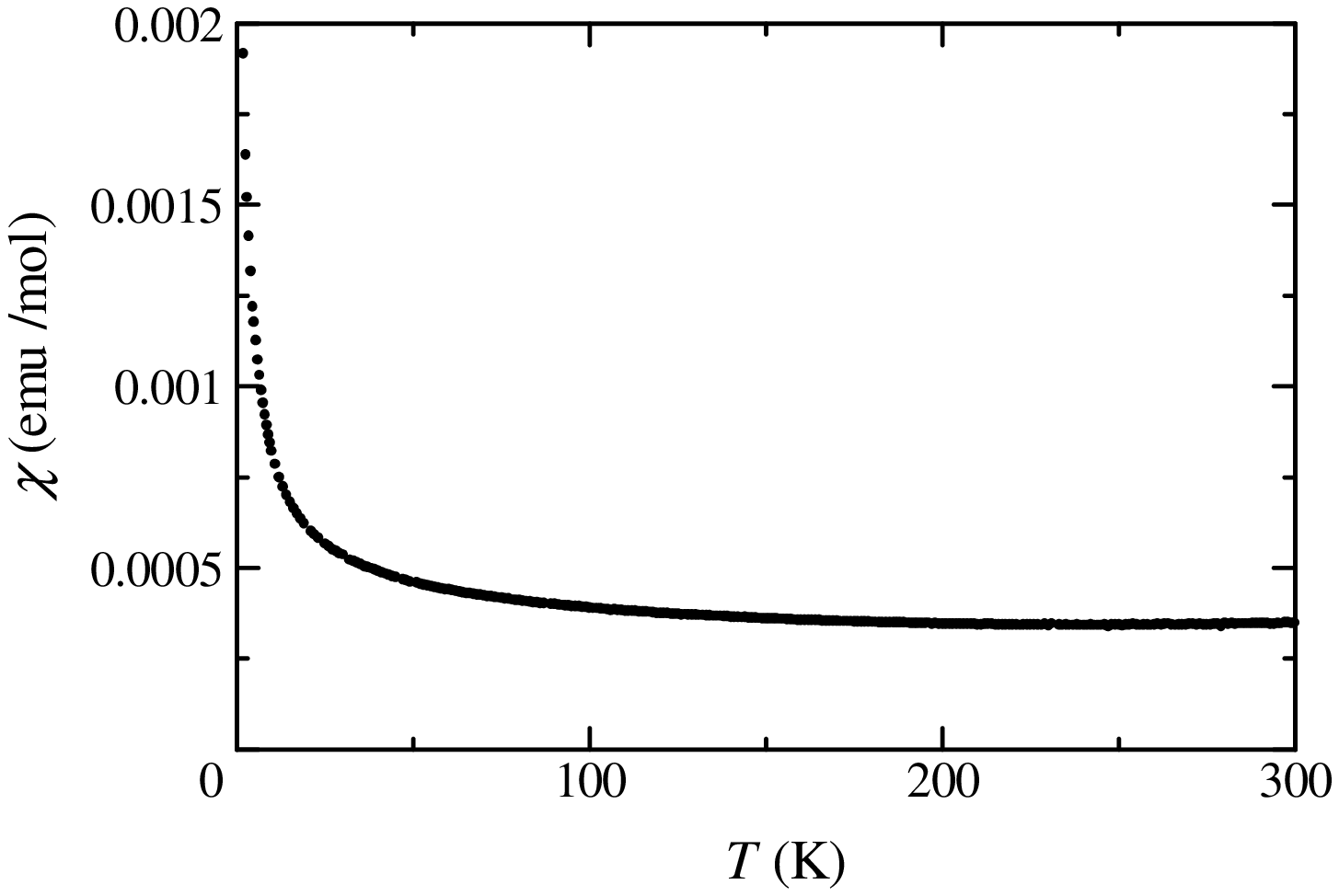}
\end{center}
\caption{Temperature dependence of the magnetic susceptibility of Li$_{2}$RuO$_{3}$ under 1 T of magnetic field.}
\label{fig7}
\end{figure}

Figure~\ref{fig7} shows the magnetic susceptibility of Li$_{2}$RuO$_{3}$ prepared by a hydrothermal method. At first, we tried to analyze the temperature dependence using Curie-Weiss formula, but it was not successful. Then, we took account of also a constant term $\chi_{0}$. Using the formula
\begin{equation}
\label{eq1}
\chi = \frac{C}{T-\Theta} + \chi_{0},
\end{equation}
the observed temperature dependence is reproduced. The values of $C$, $\Theta$ and $\chi_{0}$ obtained by a least-square fitting were $2.9 \times 10^{-3}$ emu$\cdot$K/mol, -1.9 K and $2.9 \times 10^{-4}$ emu/mol, respectively. The Curie constant is only 0.3 \% of what is expected from a localized spin model with Ru$^{4+}$ ($S = 1$).

One of the possible explanations about this discrepancy is that the $d$-electrons are delocalized in the two-dimensional network. The Curie-Weiss term is due to an extrinsic effect such as magnetic impurities or defects. In this picture, the susceptibility $\chi_{0}$, independent of temperature, can be interpreted in terms of Pauli paramagnetism. The value of $\chi_{0}$, $\approx 3 \times 10^{-4}$ emu/mol seems large for Pauli paramagnetism, but not extraordinary for highly correlated oxides. However, a preliminary study of the electric conductivity on a pelletized powder measured only $\approx 10^{-4}$ $\Omega^{-1}$cm$^{-1}$ at room temperature, which is far from a typical values in metals. For the reason, we cannot exclude the contact resistances between particles at present. 

Another interpretation is that the four $4d$ electrons in Ru$^{4+}$ form a spinless state. The triple degeneracy of the $t_{2g}$ levels is lifted due to a trigonal distortion of the RuO$_{6}$ octahadra. If the four $4d$ electrons occupy the lower two levels, the total spin degree of freedom disappears. In this case, we need another explanation about the origin of the $\chi_{0}$ term. If the energy level of the excitation state with a normal spin configuration is close to the ground state, it might contribute to the temperature independent paramagnetism through a second order parturbation.

Recently, an interesting interpretation of non-magnetic behavior of Li$_{2}$RuO$_{3}$ has been given by Miura \emph{et al.}\cite{miura06a} They pointed out that a large distortion from an ideal honeycomb lattice has fatal effect on the magnetic ground state, because the spins form singlet pairs on the shortest Ru-Ru bonds. Similar phenomena were observed in several oxides having a distorted honeycomb structure.\cite{miura06b}

\subsection{Magnetism of cubic-Li$_{x}$RuO$_{1+x}$}
\begin{figure}[tb]
\begin{center}
\includegraphics[width=1.0\linewidth]{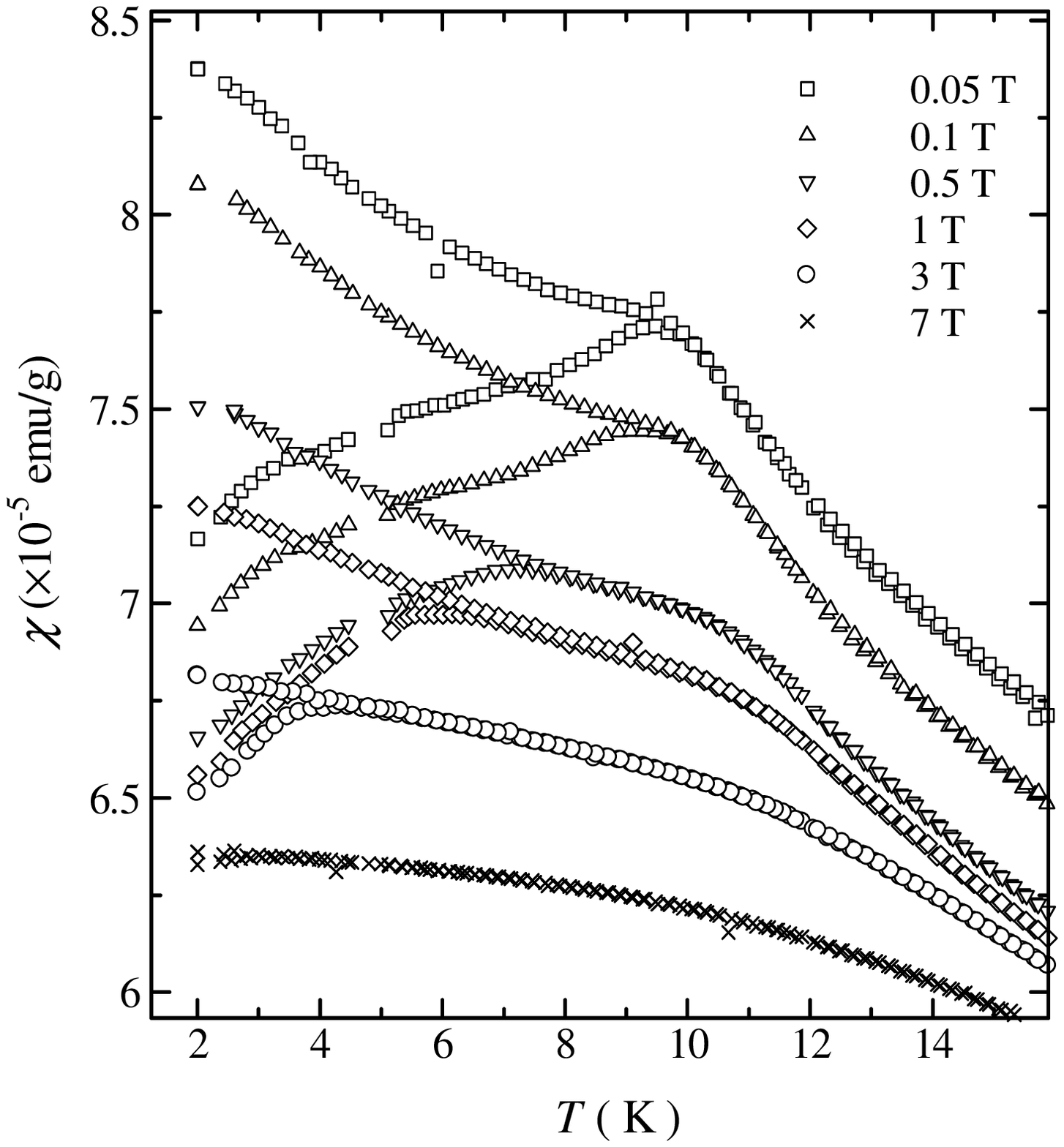}
\end{center}
\caption{Temperature dependence of the DC magnetic susceptibility of cubic-Li$_{x}$RuO$_{1+x}$ under various magnetic fields. For $B = 0.05$ T, the magnetic susceptibility exhibits a cusp at 10 K, below which irreversibility appears between FC (upper) and ZFC (lower) curves. Under higher magnetic fields, the cusp becomes less sharp and the onset temperature of the irreversibility becomes lower.}
\label{fig8}
\end{figure}

The magnetic susceptibility of cubic-Li$_{x}$RuO$_{1+x}$ above 20 K simply obeyed Curie-Weiss law. The obtained Curie constant and Weiss temperature were $C = 0.0078$ emu$\cdot$K/g and $\Theta = -100$ K, respectively. As already mentioned, the composition of this phase guessed from the X-ray diffraction is closed to ``Li$_{3}$RuO$_{4}$''. Assuming this composition, the molar Curie constant is calculated as $C = 1.4$ emu$\cdot$K/mol, which is close to the calculated value for Ru$^{5+}$, 1.875 emu$\cdot$K/mol. The Weiss temperature, $-100$ K, with a large absolute value indicates the existence of strong antiferromagnetic interactions. However, this compound does not undergo an antiferromagnetic phase transition. Instead, a spin-glass behavior appears. The magnetic susceptibility was measured under various DC magnetic fields, as shown in Fig.~\ref{fig8}. At 0.05 T, the susceptibility exhibits a cusp at 10 K, below which the ZFC curve deviates from the FC curve. As the magnetic field becomes higher, the cusp becomes less sharp and the onset temperature of the irreversiblility becomes lower. These observations agree with a behavior of an ordinary spin glass.\cite{binder86} In order to verify that the irreversibility is an intrinsic effect, we also performed a measurement on single crystals tightly fixed to the straw with grease. The result was almost the same. 

\begin{figure}[tb]
\begin{center}
\includegraphics[width=1.0\linewidth]{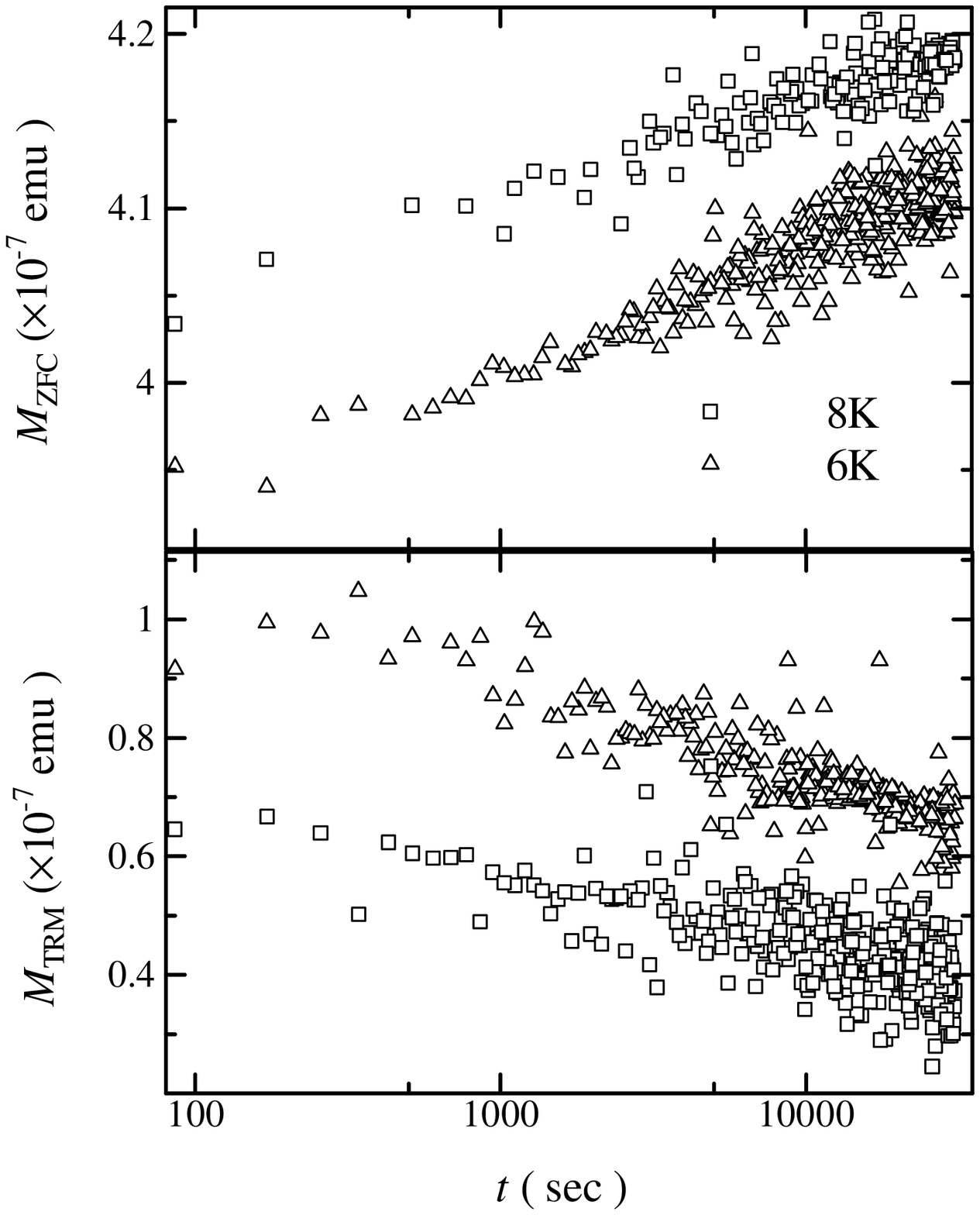}
\end{center}
\caption{Time dependence of the magnetization of cubic-Li$_{x}$RuO$_{1+x}$. In ZFC condition, the sample was cooled under zero magnetic field. After 3000 s, 3.33 mT of magnetic field was applied. In thermoremanent magnetization (TRM) condition, the sample was cooled under 100 mT of magnetic field. After 3000 s, the magnetic field was quickly removed. In both cases, aging phenomena are clearly observed.}
\label{fig9}
\end{figure}

We investigated the time dependences of the magnetization under the following conditions. In ZFC condition, the sample was cooled from 50 K to the target temperature under zero magnetic field. After waiting for 3000 s, 3.33 mT of magnetic field was applied and the time dependence of the magnetization was recorded. In thermoremanent magnetization (TRM) condition, we first cooled the sample from 50 K to the target temperature in 100 mT of magnetic field. After waiting for 3000 s, the magnetic field was quickly removed and the magnetization was recorded as a function of time. The results are shown in Fig.~\ref{fig9}. In the both conditions, aging phenomena clearly appeared; the magnetizations significantly dependent on time. Although the large experimental error prevents us from obtaining the exact function, the time dependence seems not very different from an exponential-type decay.

\begin{figure}[tb]
\begin{center}
\includegraphics[width=1.0\linewidth]{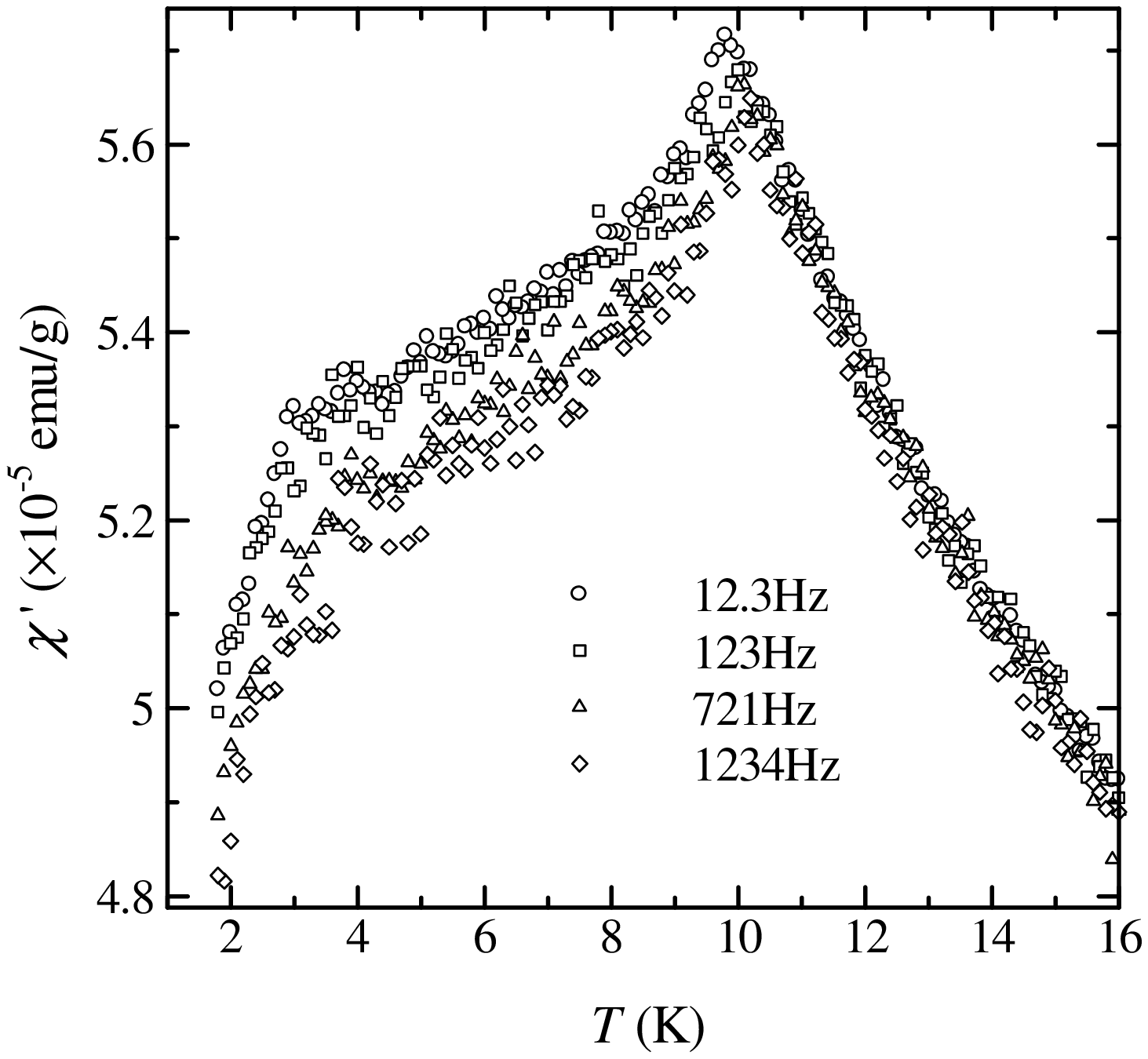}
\end{center}
\caption{Real components of the AC magnetic susceptibility of cubic-Li$_{x}$RuO$_{1+x}$ under serveral frequencies. The amplitude of the AC magnetic field was fixed to 0.65 mT. The contribution from the grease was not subtracted. That is why the valued of the susceptibility is slightly smaller than what measured by DC method. }
\label{fig10}
\end{figure}

We also measured the AC magnetic susceptibility using a SQUID magnetometer. The amplitude of the AC field was 0.65 mT. As shown in Fig.~\ref{fig10}, the real component of the AC susceptibility, $\chi'$, detected a sharp cusp at 10 K, below which the imaginary component $\chi''$ rises. For higher frequencies, the cusp temperature becomes higher. This is a common behavior in spin-glass systems. The AC susceptibility also detected a sharp bend at 3 K, suggesting another phase transition. This transition temperature also slightly shifts to higher temperature for higher frequency. This indicates that the 3 K transition is also a kind of glass transition.  

``Spin-glass'' like behaviors had been, so far, reported in several ruthenates. Among them, CaRuO$_{3}$\cite{felner00} and Y$_{2}$Ru$_{2}$O$_{7}$\cite{yoshii99,ito00} do not contain magnetic elements other than Ru. CaRuO$_{3}$ had been regard as a paramagnetic metal at first. Felner, \emph{et al}. reported that its magnetic susceptibility exhibits an irreversibility below 90 K.\cite{felner00} They mentioned that CaRuO$_{3}$ is on the verge of magnetic ordering and readily evolves into a magnetically ordered phase. CaRuO$_{3}$ is, however, unlike a spin glass in ordinary sence. Because, there is no cusp in the magnetic susceptibility at the onset temperature of irreversibility. In addition, no aging effect nor frequency-dependent magnetism were reported as far as our knowledge. Y$_{2}$Ru$_{2}$O$_{7}$ has a pyrochlore lattice with a geometric frustration. In this compound, an irreversibility appears in the magnetic susceptibility below 76.5 K. However, neutron experiments revealed sharp magnetic scatterings below 76.5 K, which indicates a development of a long-range magnetic order unlike ordinary spin-glass states.\cite{ito00} 

As we showed a number of evidences here, cubic-Li$_{x}$RuO$_{1+x}$ definitely exhibits a spin-glass behavior. Let us consider its origin. Firstly, we should point out that there is a structural disorder; Li and Ru randomly occupy the octahedral interstitial sites in the fcc lattice of oxygen. A wide distribution of the exchange interactions may suppress a development of a long-range antiferromagnetic order. In addition, a geometric frustration is inherent in the fcc lattice of cations. This prevents it from a development of a trivial N\'eel ordering even if all of the cation sites were occupied by Ru atoms. The geometry of the magnetic ions are equivalent to that in Eu$_{x}$Sr$_{1-x}$S in which magnetic Eu$^{2+}$ and nonmagnetic Sr$^{2+}$ randomly occupy the cation sites of a NaCl lattice.\cite{maletta79a,maletta79b} In that case, a transition from superparamagnetism to spin glass is observed at $x = x_{p}$ $(\approx 0.136)$, where $x_{p}$ is a percolation limit for fcc lattice taking account of the nearest and the next-nearest neighbor exchange interactions.\cite{dalton64}

Most of spin-glass systems have some anisotropy. For example, Mn$_{x}$Fe$_{1-x}$TiO$_{3}$ and Ni$_{x}$Mn$_{1-x}$TiO$_{3}$ have Ising-type and $XY$-type anisotropies, respectively.\cite{ito88,kawano93} In Heisenberg spin-glass system with a weak anisotropy, a mean-field theory predicted that the transverse component of spins freezes at higher temperature than the temperature where the longitudinal component does.\cite{gabay81} As a result, two successive phase transitions appear. Such a phenomenon was, for example, reported on ZnMn\cite{murayama86} or on Ni$_{x}$Mn$_{1-x}$TiO$_{3}$\cite{kawano93} where there is a single-ion type anisotropy. Let us consider the dimensionality and the spin degree-of-freedom of our cubic-Li$_{x}$RuO$_{1+x}$. Because each ruthenium ion is located on the center of the octahedron of oxygen ions, it is clear that there is no reason to have a single-ion type anisotropy. Furthermore, as we disscussed above, the oxidization number of Ru is 5+. In Ru$^{5+}$, each of the three $t_{2g}$ orbitals is singly occupied by an electron, and consequently no orbital degree-of-freedom remains. Therefore, we can regard this compound as a three-dimensional Heisenberg-spin system as a first approximation.

In conventional models of spin glass, it has been considered that Ising-type anisotropy is essential for a spin-glass transition at finite temperature, although a number of experiments on almost purely Heisenberg spin-glass system have revealed a glass transition at finite temperature. Recently, Kawamura has proposed another interpretation of the glass transition in Heisenberg system called ``chirality hypothesis''.\cite{kawamura92,kawamura96} In his theory, Heisenberg-type spins themselves do not freeze but chiralities undergo a glass transition at finite temperature in Heisenberg system. The anomaly of the chirality can be detected by a magnetization measurement if there is a spin-chirality coupling.\cite{imagawa04,hukushima05} Based on this model, the spin-glass transition at finite magnetic field is not a two-step type. This is in contrast with a mean-field prediction where a spin-glass transition is a single step at zero field, but becomes two steps under a finite magnetic field because of the anistropy introduced by an external field.\cite{gabay81,almeida78}

Unlike another Heisenberg-glass Y$_{2}$Mo$_{2}$O$_{7}$,\cite{miyoshi00} the spin-glass transition in cubic-Li$_{x}$RuO$_{1+x}$ around at 10 K seems always single-step under high magnetic fields, as far as we look at the behavior of the irreversibility. This does not conflict with the chirality hypothesis. On the other hand, the AC suscepitibility detected another glass transition at 3 K, suggesting a multi-phase character of the glass state in cubic-Li$_{x}$RuO$_{1+x}$. More investigations are necessary to clarify its origin. As a possible interpretation, a conventional spin-glass transition due to a very weak anisotropy might occure at 3 K, in addition to a chiral-glass transition at 10 K. 

\section{Conclusion}
In order to compare magnetisms with various networks of exchange interactions, we synthesized several types of NaCl-type lithium ruthenates by a hydrothermal technique. Li$_{3}$RuO$_{4}$ has a structure characterized by one-dimensional zigzag chains of Ru. Its magnetic susceptibility exhibits successive phase transitions: an antiferromagnetic transition at 66 K and another transition at 32 K with an irreversibility between ZFC and FC magnetic susceptibilities. Li$_{2}$RuO$_{3}$ comprising a two-dimensional honeycomb network of Ru has a magnetic susceptibility almost independent of temperature. We discovered a novel lithium ruthenate, cubic-Li$_{x}$RuO$_{1+x}$, in which Ru and Li randomly occupy the octahedral interstitial sites of the fcc lattice of oxygen. The magnetic susceptibility exhibits a cusp at 10 K below which an irreversibility appears between the ZFC and the FC curves, characterizing a spin glass. Aging phenomena of magnetization and a frequency-dependent AC susceptibility also evidenced a spin-glass state. AC susceptibility revealed another phase transition at 3 K, suggesting a multi-phase character of the glass state.

\section*{Acknowledgment}
We would like to thank Dr. K. Miyagawa and Prof. K. Kanoda for giving us valuable opportunity for using their SQUID for AC susceptibility measurments. We also thank Prof. K. Ohishi for AC specific heat measurements. This study was supported by Grants-in-Aid for Scientific Research No. 15750127 and No. 17540341 from the Ministry of Education, Culture, Sports, Science and Technology.


\begin{thebibliography}{99} 

\bibitem{randall59} J.J. Randall and R. Ward: J. Am. Chem. Soc. \textbf{81} (1959) 2629.
\bibitem{cao97} G. Cao, S. McCall, M. Shepard, J.E. Crow and R.P. Guertin: Phys. Rev. B \textbf{56} (1997) 321.

\bibitem{maeno94} Y. Maeno, H. Hashimoto, K. Yoshida, S. Nishizaki, T. Fujita, J.G. Bednorz and F. Lichtenberg: Nature \textbf{372} (1994) 532.
\bibitem{ishida98} K. Ishida, H. Mukuda, Y. Kitaoka, K. Asayama, Z.Q. Mao, Y. Mori and Y. Maeno: Nature \textbf{396} (1998) 658.

\bibitem{schollwock04} U. Schollw\"{o}ck, J. Richter, D.J.J. Farnell and R.F. Bishop (Eds.): \textit{Quantum Magnetism} (Springer, Berlin, 2004).

\bibitem{binder86} K. Binder and A.P. Young: Rev. Mod. Phys. \textbf{58} (1986) 801.

\bibitem{alexander03} A. Alexander, P.D. Battle, J.C. Burley, D.J. Gallon, C.P. Grey and S.H. Kim: J. Mater. Chem. \textbf{13} (2003) 2612.

\bibitem{kobayashi95} H. Kobayashi, R. Kanno, Y. Kawamoto, M. Tabuchi, O. Nakamura and M. Takano: Solid State Ionics \textbf{82} (1995) 25.
\bibitem{james88} A.C.W.P. James and J.B. Goodenough: J. Solid State Chem. \textbf{74} (1988) 287.

\bibitem{izumi00} F. Izumi and T. Ikeda: Mater. Sci. Forum 198 (2000) 321.

\bibitem{anderson64} J.C. Anderson and M. Schieber: J. Phys. Chem. Solid \textbf{25} (1964) 961.

\bibitem{bonner64} J.C. Bonner and M.E. Fisher: Phys. Rev. \textbf{135} (1964) A640.

\bibitem{ito88} A. Ito, H. Aruga, M. Kikuchi, Y. Syono and H. Takei: Solid State Commun. \textbf{66} (1988) 475.

\bibitem{miura06a} Y. Miura, Y. Yasui, M. Sato, N. Igawa and K. Kakurai: Meeting Abst. of the Phys. Soc. Jpn. \textbf{61} (2006) 473.
\bibitem{miura06b} Y. Miura, R. Hirai, Y. Kobayashi and M. Sato: J. Phys. Soc. Jpn. \textbf{75} (2006) 084707.

\bibitem{felner00} I. Felner, I. Nowik, I. Bradaric and M. Gospodinov: Phys. Rev. B \textbf{62} (2000) 11332.

\bibitem{ito00} M. Ito, Y. Yasui, M. Kanada, H. Harashina, S. Yoshii, K. Murata, M. Sato, H. Okumura and K. Kakurai: J. Phys. Soc. Jpn. \textbf{69} (2000) 888.
\bibitem{yoshii99} S. Yoshii and M. Sato: J. Phys. Soc. Jpn. \textbf{68} (1999) 3034.

\bibitem{maletta79a} H. Maletta and P. Convert: Phys. Rev. Lett. \textbf{42} (1979) 108.
\bibitem{maletta79b} H. Maletta and W. Felsch: Phy. Rev. B \textbf{20} (1979) 1245.
\bibitem{dalton64} N.W. Dalton, C. Domb and M.F. Sykes, Proc. Phys. Soc. London \textbf{83} (1964) 496.

\bibitem{kawano93} H. Kawano, H. Yoshizawa, A. Ito and K. Motoya: J. Phys. Soc. Jpn. \textbf{62} (1993) 2575.

\bibitem{gabay81} M. Gabay and G. Toulouse: Phys. Rev. Lett. \textbf{47} (1981) 201.

\bibitem{murayama86} S. Murayama, K. Yokosawa, Y. Miyako and E.F. Wassermann: Phys. Rev. Lett. \textbf{57} (1986) 1785.

\bibitem{kawamura92} H. Kawamura: Phys. Rev. Lett. \textbf{68} (1992) 3785.
\bibitem{kawamura96} H. Kawamura: Int. J. Mod. Phys. \textbf{7} (1996) 345.

\bibitem{imagawa04} D. Imagawa and H. Kawamura: Phys. Rev. Lett. \textbf{92} (2004) 077204.

\bibitem{hukushima05} K. Hukushima and H. Kawamura: Phys. Rev. B \textbf{72} (2005) 144416.

\bibitem{almeida78} J.R.L. de Almeida and D.J. Thouless: J. Phys. A \textbf{11} (1978) 983.

\bibitem{miyoshi00} K. Miyoshi, Y. Nishimura, K. Honda, K. Fujiwara and J. Takeuchi: J. Phys. Soc. Jpn. \textbf{69} (2000) 3517.

\end{thebibliography}
\end{document}